\documentclass[aps,prd,twocolumn]{revtex4}
\begin{document}
\title{Thermodynamic pressure for massless QCD and the trace anomaly}
\author{H. Arthur Weldon}
\email[]{hweldon@WVU.edu}
\affiliation{Department of Physics and Astronomy, West Virginia University , Morgantown, West Virginia, 26506-6315}
\date{\today}

\begin{abstract}
From statistical mechanics the trace of the thermal average of any energy-momentum tensor is
$\langle T^{\mu}_{\;\;\mu}\rangle =T\partial P/\partial T-4P$. The renormalization group  formula
$\langle T^{\mu}_{\;\;\mu}\rangle =\beta(g_{M})\partial P/\partial g_{M}$ for QCD with massless fermions requires the pressure to
have the structure
\begin{displaymath} P=T^{4}\sum_{n=0}^{\infty} \phi_{n}(g_{M})\big[\ln\big({M\over 4\pi T}\big)\big]^{n},\end{displaymath}
where the factor $4\pi$ is for later convenience.  The functions $\phi_{n}(g_{M})$ for $n\ge 1$ may be calculated from
$\phi_{0}(g_{M})$ using the recursion relation $n\,\phi_{n}(g_{M})=-\beta(g_{M})d\phi_{n-1}/dg_{M}$.  This is checked 
against known perturbation theory results by using the terms of order $(g_{M})^{2}, (g_{M})^{3}$, $(g_{M})^{4}$ in $\phi_{0}(g_{M})$
to obtain the known terms of order $(g_{M})^{4}, (g_{M})^{5}$,  $(g_{M})^{6}$ in $\phi_{1}(g_{M})$ and the known term of order
$(g_{M})^{6}$ in $\phi_{2}(g_{M})$. The above series may be summed and gives the same result as choosing $M=4\pi T$, viz.  
$T^{4}\phi_{0}(g_{4\pi T})$. \end{abstract}

\maketitle

\section{Introduction}

For a symmetric energy-momentum tensor $T^{\mu\nu}$ the dilation current $S^{\mu}=T^{\mu\lambda}x_{\lambda}$ and the four
conformal currents $K^{\alpha\mu}=x^{2}T^{\alpha\mu}-2x^{\alpha}T^{\mu\lambda}x_{\lambda}$ are conserved if the energy-momentum
tensor is traceless:
\begin{eqnarray}
\partial_{\mu}S^{\mu}&=& T^{\mu}_{\;\;\mu}\nonumber\\
\partial_{\mu}K^{\alpha\mu}&=&-2x^{\alpha}T^{\mu}_{\;\;\mu}.\nonumber\end{eqnarray}
The classical energy-momentum tensor for QCD with massless fermions is traceless but quantum corrections introduce
a renormalization scale that spoils the conservation of scale and conformal currents and renders the
trace nonzero \cite{Coleman}.

The trace of the thermally averaged energy-momentum tensor is
$\langle T^{\mu}_{\;\;\mu}\rangle =u-3P$ where $u=\langle T^{0}_{\;\;0}\rangle$ is the energy density
and $P=-\sum_{j=1}^{3}\langle T^{j}_{\;\;j}\rangle/3$ is the pressure.  The relation 
\begin{displaymath}
\exp(\beta PV)=Z={\rm Tr}\{e^{-\beta H}\}\end{displaymath}
between the pressure and the
partition function implies that 
\begin{displaymath}
{\partial\over\partial\beta} (\beta P)=-{\langle H\rangle\over V}=-u,\end{displaymath}
or equivalently
\begin{displaymath}
T{\partial P\over\partial T}=u+P.\end{displaymath}
The trace of the energy-momentum tensor becomes
\begin{equation}
\langle T^{\mu}_{\;\;\mu}\rangle=u-3P=T{\partial P\over \partial T}-4P.\label{one}\end{equation}

For non-Abelian gauge fields with massless fermions the pressure has the form
\begin{equation}
P=T^{4}\Phi(g_{M}, M/T),\label{Phi}\end{equation}
where $M$ is the renormalization scale. From (\ref{one})  the trace of the energy-momentum tensor is
\begin{equation}
\langle T^{\mu}_{\;\;\mu}\rangle=T^{5}{\partial\Phi\over\partial T}.\label{statmech} \end{equation}
One would expect the calculation of $\Phi$ to be primary and the trace anomaly only an afterthought.
However with the theorem of Drummond, Horgan, Landshoff, and Rebhan \cite{Landshoff} that
\begin{equation}
\langle T^{\mu}_{\;\;\mu}\rangle=\beta(g_{M}){\partial P\over \partial g_{M}}\label{Landshoff}\end{equation}
the anomaly becomes predictive in that the combination of (\ref{statmech}) and (\ref{Landshoff}) gives
\begin{equation}
T{\partial\Phi\over\partial T}=\beta(g_{M}){\partial\Phi\over \partial g_{M}},\label{identity}\end{equation}
which is Eq. (3.11) of Drummond et al \cite{Landshoff}.

Note that (\ref{Landshoff}) is similar to the zero temperature operator identity 
$T^{\mu}_{\;\;\mu}=\beta(g_{M})\partial{\cal L}/\partial g_{M}$.

Sec. II shows how Eq. (\ref{identity}) ensures that $P$ is independent of the renormalization scale $M$ and requires $P$ to have the
structure shown in the abstract. Sec. III tests the recursion relation using known results for $\phi_{0}(g_{M})$ from perturbation theory
to calculate the three known terms in $\phi_{1}(g_{M})$ and the only known term of $\phi_{2}(g_{M})$ and illustrates how to
improve perturbation theory.  

\section{Structure of P}

\subsubsection{Independence of the renormalization scale $M$}

As indicated in Eq (\ref{Phi}) the renormalization scale appears in $\Phi$ through $g_{M}$ 
and through $r=M/T$.  The full $M$ derivative of $\Phi$ is
\begin{equation}
M{d\Phi\over dM}=M{dg_{M}\over dM}{\partial\Phi\over\partial g_{M}}\bigg|_{r}+M{dr\over dM}{\partial\Phi\over\partial r}\bigg|_{g_{M}}.
\end{equation}
In the first term use $Mdg_{M}/dM=\beta(g_{M})$; in the second, $Mdr/dM=r$ and $r\partial\Phi/\partial r =-T\partial
\Phi/\partial T$ so that
\begin{equation} M{d\Phi\over dM}= \beta(g_{M}){\partial\Phi\over\partial g_{M}}\bigg|_{r}-T{\partial \Phi\over \partial T}\bigg|_{g_{M}}=0
\end{equation}
after using Eq (\ref{identity}).

\underbar{Comment}: One can reverse the argument and derive the anomaly relation (\ref{Landshoff}) of Drummond et al \cite{Landshoff} by starting with the assertion that $P$ is a physical quantity and must therefore be independent of the renormalization scale. 

\subsubsection{Origin of $\;[\ln(M/T)]^{n}$}

Since $\Phi(g_{M}, M/T)$ is independent of $M$ it must be only a function of $ T/\Lambda_{QCD}$.
It is convenient to consider $\Phi$ as a function $\phi_{0}$  of $\ln(\xi T/\Lambda_{QCD})$, where $\xi$ is some constant
\begin{equation}
\Phi(g_{M}, T/M)=\phi_{0}(\ln(\xi T/\Lambda_{QCD})),\end{equation}  
and to introduce variables
\begin{eqnarray}
u&=&\ln(M/\Lambda_{QCD})\cr
v&=&\ln(M/\xi T).
\end{eqnarray}
 The running coupling is a function of $u$
 determined by $\beta(g_{M})=dg_{M}/du$; $\Phi$ is a function of 
$u-v$:
\begin{eqnarray}
\Phi(g_{M},M/T)&=& \phi_{0}(u-v)\nonumber\\
&=& \sum_{n=0}^{\infty}{(-1)^{n}\over n!} {d^{n}\phi_{0}(u)\over du^{n}} v^{n},\label{Phi series}
\end{eqnarray}
after at Taylor series expansion. The definition 
\begin{equation}
\phi_{n}(g_{M})={(-1)^{n}\over n!} {d^{n}\phi_{0}(g_{M})\over du^{n}}\label{dphi/du}\end{equation}
allows the series to be written
\begin{equation}
\Phi(g_{M},M/T)=\sum_{n=0}^{\infty} \phi_{n}(g_{M})\bigg[\ln\Big({M\over\xi T}\Big)\bigg]^{n}.\label{series}\end{equation}
The recursion relation $n\,\phi_{n}(g_{M})=-d \phi_{n-1}/du$, which follows from (\ref{dphi/du}), may be expressed as
\begin{equation}
\phi_{n}(g_{M})=-{1\over n}\beta(g_{M}){d\phi_{n-1}\over dg_{M}}\hskip0.5cm (n\ge 1).\label{recursion}\end{equation}
One can confirm directly that the series (\ref{series}) satisfies $d\Phi/dM=0$.

\underbar{Comment}: If $\xi$ is changed to $\xi'$ then
\begin{equation}
\ln\Big({M\over\xi T}\Big)=\ln\Big({M\over \xi' T}\Big)+\ln\Big({\xi'\over\xi}\Big).\end{equation}
The binomial theorem allow the series (\ref{series}) to be expressed in terms of powers of $\ln(M/\xi' T)$ with modified functions
$\phi'_{n}(g_{M})$.

\underbar{Comment}: From $u-3P=T^{5}{\partial\Phi/\partial T}$ it follows that the energy density and entropy density are
\begin{eqnarray} u&=& T^{4}\big[3\Phi+T{\partial\Phi\over\partial T}]\\
s&=& T^{3}[4\Phi+T{\partial\Phi\over\partial T}\big].\end{eqnarray}

\section{Results from perturbation theory}

The ${\cal O}(g_{M}^{2})$ term in $P$ was calculated by Shuryak \cite{g2}; the ${\cal O}(g_{M}^{3})$ term by Kapusta \cite{g3};
to this order there was no $\ln(M/T)$.  The ${\cal O}(g_{M}^{4})$ term was calculated by Arnold and Zhai \cite{g4}; the ${\cal O}(g_{M}^{5})$ by Zhai and Kastening \cite{g5}; in both cases $\ln(M/T)$ appeared. The same result was obtained by
Braaten and Nieto \cite{Eric} using hard thermal loop resummation.

At  ${\cal O}(g_{M}^{6})$ nonperturbative  magnetic screening effects arise \cite{Linde,GPY,Landshoff2}.
 Kajantie et al \cite{Kajantie} were able to calculate the ${\cal O}(g_{M}^{6})$ perturbative terms and found both $\ln(M/T)$ and $\ln^{2}(M/T)$.  A convenient reference that discusses all the results is Sec. 8.4 of Kapusta and Gale \cite{Kapusta}.

\subsection{Checks against known results}

For comparison with the published results from perturbation theory it is convenient to insert a prefactor in the the series expression for the pressure and choose $\xi=4\pi$:
\begin{equation}
P={\pi^{2}d_{A}\over 9}T^{4} \sum_{n=0}^{\infty}\phi_{n}(g_{M})\bigg[\ln\bigg({M\over 4\pi T}\bigg)\bigg]^{n},\label{PIII}\end{equation}
where $d_{A}$ is the dimension of the adjoint representation.

With the order $(g_{M})^{2},(g_{M})^{3}$, and $(g_{M})^{4}$ terms of  $\phi_{0}(g_{M})$ the recursion relation (\ref{recursion}) gives the first three
terms of $\phi_{1}(g_{M})$ and the first term of $\phi_{2}(g_{M})$. 
Using the notation $\phi_{n}^{(k)}(g_{M})$ for the ${\cal O}(g_{M})^{k}$ term in $\phi_{n}(g_{M})$ the necessary inputs are
\begin{eqnarray}
\phi_{0}^{(2)}(g_{M})&=&-\Big({g_{M}\over 4\pi}\Big)^{2}(C_{A}+{5\over 2}S_{F})\cr
\phi_{0}^{(3)}(g_{M})&=&\Big({g_{M}\over 4\pi}\Big)^{3}(C_{A}+S_{F})^{3/2}16/\sqrt{3}\cr
\phi_{0}^{(4)}(g_{M})&=&\Big({g_{M}\over 4\pi}\Big)^{4}
\Big\{48 C_{A}(C_{A}\!+\!S_{F})\ln W+R\},\nonumber
\end{eqnarray}
where $W=(g_{M}/2\pi)\sqrt{(C_{A}+S_{F})/3}$ and 
\begin{equation}
R=C_{A}^{2}R_{1}+C_{A}S_{F}R_{2}+S_{F}^{2}R_{3}+S_{2F}R_{4}.\end{equation}
The coefficients $R_{j}$ are given in \cite{g4,Kapusta} in terms of Riemann zeta functions and the Euler constant. For later comparison with 
\cite{Kajantie} it is convenient to employ the approximate numerical values:
\begin{equation}\begin{array}{ll}
R_{1}=79.2626 & R_{2}=18.9212\cr
R_{3}=-0.6914 & R_{4}=9.6145.\end{array}\end{equation}
The standard notation  \cite{Kapusta}  for  SU(N) with $n_{f}$  fermions in the fundamental representation is $d_{A}\!=\!N^{2}-1,
C_{A}\!=\!N, d_{F}=Nn_{f}, S_{F}=n_{f}/2, S_{2F}=(N^{2}-1)n_{f}/4N$.
The first two terms in the beta function are
\begin{eqnarray}
\beta(g_{M})&=&-\beta_{0}g_{M}^{3}-\beta_{1}g_{M}^{5}+\dots\cr\cr
\beta_{0}&=&\Big({11\over 3}C_{A}-{4\over 3} S_{F}\Big)/(4\pi)^{2}\cr
\cr
\beta_{1}&=& \Big({34\over 3} C^{2}_{A}-{20\over 3}C_{A}S_{F}-4S_{2F}\Big)/(4\pi)^{4}.
\end{eqnarray}
The predictions of the recursion relation (\ref{recursion}) are
 \begin{eqnarray}
A.\hskip0.5cm \phi_{1}^{(4)}(g_{M})&=&\beta_{0}g_{M}^{3} {d\over dg_{M}} \phi_{0}^{(2)}(g_{M})\cr
B.\hskip0.5cm \phi_{1}^{(5)}(g_{M})&=&\beta_{0}g_{M}^{3}{d\over dg_{M}}\phi_{0}^{(3)}(g_{M}) \cr
C.\hskip0.5cm \phi_{1}^{(6)}(g_{M})&=&\beta_{0}g_{M}^{3}{d\over dg_{M}}\phi_{0}^{(4)}(g_{M})\cr
&+&\beta_{1}g_{M}^{5}{d\over dg_{M}}\phi_{0}^{(2)}(g_{M}) \cr
D.\hskip0.5cm \phi_{2}^{(6)}(g_{M})&=& {1\over 2}\beta_{0}g_{M}^{3}{d\over dg_{M}}\phi_{1}^{(4)}(g_{M}).
\end{eqnarray}

The result for A,
\begin{equation}
\phi_{1}^{(4)}(g_{M})=\Big({g_{M}\over 4\pi}\Big)^{4}\Big\{-C_{A}^{2}{22\over 3}-C_{A}S_{F}{47\over 3}+S_{F}^{2}{20\over 3}\Big\},\end{equation}
agrees with  \cite{g4,g5,Eric,Kajantie}. 

The result for B, 
\begin{eqnarray}
\phi_{1}^{(5)}(g_{M})&=&\Big({g_{M}\over 4\pi}\Big)^{5}\Big({C_{A}+S_{F}\over 3}\Big)^{1/2}\cr
\cr
&&\hskip0.5cm \times \Big(C_{A}^{2}176 +C_{A}S_{F}112
-S_{F}^{2}64\Big),\end{eqnarray}
agrees with \cite{g5,Eric,Kajantie}.

The result for  C is 
\begin{eqnarray} 
\phi_{1}^{(6)}(g_{M})&=&4\Big({g_{M}\over 4\pi}\Big)^{6}\Big\{\big({11\over 3} C_{A}\!-\!{4\over 3}S_{F}\big)R \label{phi6}\cr\cr
&&\hskip-1cm +\big( C_{A}\!+\!{5\over 2}S_{F}\big) \big(\!-{17\over 3}C^{2}_{A}\!+\!{10\over 3}C_{A}S_{F}\!+\!2S_{2F}\big) \cr\cr
&&\hskip-1.3cm +\big({11\over 3}C_{A}\!-\!{4\over 3}S_{F}\big)C_{A}(C_{A}\!+\!S_{F})\big(12\!+\!48\ln W\big)\}.
\end{eqnarray}
 To compare this with \cite{Kajantie} 
it is necessary to evaluate (\ref{phi6}) for  SU(3):
\begin{eqnarray}
\phi_{1}^{(6)}(g_{M})&=&4\Big({g_{M}\over 4\pi}\Big)^{6}\Big\{\! 432\big(11\!-\!{2\over 3}n_{f}\big)\big(1\!+\!{1\over 6}n_{f}\big)\ln W\cr\cr
&&\hskip-1.5cm +1035\!+{325\over 4} n_{f} \!-{49\over 12} n_{f}^{2}\! +\!\big(11\!-\!{2\over 3}n_{f}\big)R\Big\}.
\end{eqnarray}
Substituting the numerical values of $R$ gives the final result
\begin{eqnarray}
\phi_{1}^{(6)}(g_{M})&=&4\Big({g_{M}\over 4\pi}\Big)^{6}\Big\{\! 432\big(11\!-\!{2\over 3}n_{f}\big)\big(1\!+\!{1\over 6}n_{f}\big)\ln W\cr\cr
&&\hskip-1.5 cm +8882\!-\!11.6186  n_{f} \!-\! 29.1767 n_{f}^{2}\! +\!  0.1152 n_{f}^{3}\Big\}.\label{333}
\end{eqnarray}
In \cite{Kajantie} the ${\cal O}(g_{M}^{6})$ results are expressed in terms of $(\alpha_{M}/\pi)^{3}$ and $\ln(M/2\pi T)$. When \cite{Kajantie} is reexpressed in terms of $(g_{M}/4\pi)^{6}$ and $\ln(M/4\pi T)$ it agrees completely  with Eq. (\ref{333}). 

The final calculation D gives
\begin{equation}
\phi_{2}^{(6)}(g_{M})=-\Big({g_{M}\over 4\pi}\Big)^{6}4\big(C_{A}+{5\over 2}S_{F}\big)\big({11\over 3}C_{A}-{4\over 3}S_{F}\big)^{2}.
\end{equation}
For SU(3) with $n_{f}$ multiplets of fermions
\begin{equation}
\phi_{2}^{(6)}(g_{M})=-\Big({g_{M}\over 4\pi}\Big)^{6}1452 \big(1+{5\over 12}n_{f}\big)(1-{2\over 33}n_{f}\big)^{2},
\end{equation}
which is exactly the same as \cite{Kajantie}.

\subsection{Improving perturbation theory}

 At order $(g_{M})^{6}$  nonperturbative effects appear in $\phi_{0}^{(6)}(g_{M})$ but not in $\phi_{1}^{(6)}(g_{M})$ or $\phi_{2}^{(6)}(g_{M})$ calculated above. The argument of Linde \cite{Linde,GPY,Kapusta} shows that certain diagrams that appear to be of order
$(g_{M})^{k}$ with $k>6$ are so infrared sensitive that nonperturbative magnetic shielding will render them of order $(g_{M})^{6}$.
Thus $\phi_{0}^{(6)}(g_{M})$ receives contributions from diagrams with infinitely many loops. 
Nevertheless  $\phi_{0}(g_{M})$ is still a series of the form
\begin{equation}
\phi_{0}(g_{M})=\sum_{k=0}^{\infty} \phi_{0}^{(k)}(g_{M}).\label{phi0sum}\end{equation}
The $k=1$ term vanishes; the $k=2$ term is the first to depend on $g_{M}$. 
Because the beta function begins with $(g_{M})^{3}$ the recursion relation (\ref{recursion}) implies that 
$\phi_{0}^{(k)}(g_{M})$ will generate terms of order $(g_{M})^{2n+k}[\ln(M/4\pi T)]^{n}$.
The series (\ref{PIII}) for $P$ may be considered a double series:
\begin{equation}
P={\pi^{2}d_{A}\over 9}T^{4}\sum_{k=0}^{\infty} \sum_{n=0}^{\infty}\phi^{(2n+k)}_{n}(g_{M})\bigg[\ln\bigg({M\over 4\pi T}\bigg)\bigg]^{n}.\label{Pdoublesum}\end{equation}
Perturbative calculations  through order $(g_{M})^{5}$ determine $\phi_{n}^{(2n+k)}(g_{M})$ for $2n+k\le 5$:
\begin{equation}
P^{(k\le 5)}_{[n]}={\pi^{2}d_{A}\over 9}T^{4}\sum_{k=0}^{5} \sum_{n=0}^{{1\over 2}(5-k)}\phi^{(2n+k)}_{n}(g_{M})\bigg[\ln\bigg({M\over 4\pi T}\bigg)\bigg]^{n}.\end{equation}
The difference between $P^{(k\le 5)}_{[n]}$ and $P^{k\le 4}_{[n]}$ is not small \cite{g5,Eric,Eric2}.

There is no need to terminate the sum over $n$; one can easily compute the full sum
\begin{equation}
P^{(k\le 5)}={\pi^{2}d_{A}\over 9}T^{4}\sum_{k=0}^{5} \sum_{n=0}^{\infty}\phi^{(2n+k)}_{n}(g_{M})\bigg[\ln\bigg({M\over 4\pi T}\bigg)\bigg]^{n}.\end{equation}
The input is of the form
\begin{equation}
\phi_{0}^{(k)}(g_{M})= \big({g_{M}\over 4\pi}\Big)^{k}\Big\{ A_{k}+B_{k}\ln\Big[{g_{M}\over 2\pi}\sqrt{(C_{A}\!+\!S_{F})/3}\Big]\Big\},\end{equation}
where $A_{1}=0$ and $B_{4}$ is the only nonzero $B_{k}$ for $k\le 5$.
As before, define $u=\ln(M/\Lambda_{QCD})$. At large $M$, one can use $(g_{M})^{2}=[\beta_{0}u]^{-1}$
and the parametrization
\begin{equation}\phi_{0}^{(k)}(g_{M})={1\over u^{k/2}}\big( a_{k}+b_{k}\ln u\big).\label{orderk}
\end{equation}
The n'th order derivatives of $\phi_{0}(g_{M})$ required by Eq. (\ref{dphi/du}) give
\begin{eqnarray}\phi_{n}^{(2n+k)}(g_{M})&=&{1\over u^{k/2+n}}\Big[a_{k}S_{n}-2{dS_{n}\over dk}b_{k}
 +S_{n}b_{k}\ln u\Big]\cr\cr
S_{n}&=&{\Gamma(n+k/2)\over  n!\Gamma(k/2)}.\label{Sn}\end{eqnarray}
With $v=\ln(M/4\pi T)$ Eq.  (\ref{Phi series}) requires the sum
\begin{equation}
\sum_{n=0}^{\infty}S_{n}\Big({v\over u}\Big)^{n}.
\end{equation}
By the ratio test this sum converges for $|v/u|<1$, which is satisfied provided $M>\sqrt{4\pi T\Lambda_{QCD}}$ and $4\pi T>\Lambda_{QCD}$.  The result is
\begin{equation}
\sum_{n=0}^{\infty}S_{n}\Big({v\over u}\Big)^{n}=\Big[1-{v\over u}\Big]^{-k/2}
\end{equation}
Applying $d/dk$ as required in (\ref{Sn}) gives 
\begin{eqnarray}
P^{(k\le 5)}= {\pi^{2}d_{A}\over 9}T^{4}\sum_{k=0}^{5}{1\over (u-v)^{k/2}}
\Big[a_{k}+b_{k}\ln(u-v)\Big].
\end{eqnarray}
The dependence on the renormalization scale $M$ disappears since  $u-v=\ln(4\pi T/\Lambda_{QCD})$.
When $a_{k}, b_{k}$ are expressed  in terms of $A_{k},  B_{k}$ and $u-v=(\beta_{0}g_{4\pi T})^{-1}$
the result is
\begin{eqnarray}
P^{(k\le 5)}&= &{\pi^{2}d_{A}\over 9}T^{4}\sum_{k=0}^{5}\Big({g_{4\pi T}\over 4\pi}\Big)^{k}
\Big\{A_{k}\cr
&&\hskip0.5cm+B_{k}\ln\Big[{g_{4\pi T}\over 2\pi}\sqrt{(C_{A}\!+\!S_{F})/3}\Big]\Big\};
\end{eqnarray}
or more concisely
\begin{equation}
P^{(k\le 5)}= {\pi^{2}d_{A}\over 9}T^{4}\sum_{k=0}^{5}\phi_{0}^{(k)}(g_{\mu})\Big|_{\mu=4\pi T}.
\end{equation}
In short, convergence of the infinite sum on $n$ in (\ref{Pdoublesum}) is automatic; whether a finite number of $\phi_{0}^{(k)}(g_{M})$   in the series for 
 (\ref{phi0sum}) for $\phi_{0}(g_{M})$ is a good approximation, i.e.whether perturbation theory is reliable, is an open question  \cite{Eric2}.


\begin{thebibliography}{}

\bibitem{Coleman} S. Coleman and R. Jackiw, Why Dilation Generators Do Not Generate Dilations, Ann. Phys. {\bf 67}, 552 (1971).

\bibitem{Landshoff} I.T. Drummond, R.R. Horgan, P.V.  Landshoff, and A. Rebhan, QCD Pressure and the trace anomaly,
Phys. Lett.  B {\bf 460}, 197 (1999). 

\bibitem{g2} E.V. Shuryak, ``Quark-gluon plasma and hadronic production of leptons, photons, and psions,"
 Phys. Lett. {\bf 78B}, 150 (1978).

\bibitem{g3} J.I. Kapusta, Quantum Chromodynamics at High Temperature, Nucl. Phys.  {\bf B148}, 461 (1979).

\bibitem{g4} P. Arnold and C. Zhai, Three-loop free energy for pure gauge QCD, Phys. Rev. D {\bf 50}, 7603 (1994); Three-loop free energy for high-temperature QED and QCD with fermions, Phys. Rev. D {\bf 51}, 1906 (1995).

\bibitem{g5} C. Zhai and B. Kastening, Free energy of hot gauge theories with fermions through $g^{5}$, Phys. Rev. D {\bf 52} , 7232 (1995).

\bibitem{Eric} E. Braaten and A. Nieto, On the Convergence of Perturbative QCD at High Temperature, Phys. Rev. Lett.
{\bf 76}, 1417 (1996); Free energy of QCD at high temperature, Phys. Rev. D {\bf 53}, 3421 (1996).

\bibitem{Linde} A.D. Linde, Infrared problems in the thermodynamics of the Yang-Mills gas, Phys. Lett.  {\bf 96B}, 289 (1980).

\bibitem{GPY} D.J.  Gross, R.D.  Pisarski, and L.G. Yaffe, QCD and instantons at finite temperature,  Rev. Mod. Phys. {\bf 53}, 43 (1981).

\bibitem{Landshoff2} I.T. Drummond, R.R. Horgan, P.V. Landshoff, and A. Rebhan, Eliminating infrared divergences in the pressure,
Phys. Lett. B. {\bf 398}, 326 (1997).

\bibitem{Kajantie}  K. Kajantie, M. Laine, K. Rummukainen,  Y. Schr\"oder, Pressure of hot QCD up to $g^{6}\ln(1/g)$, Phys. Rev. D
{\bf 67}, 105008 (2003).

\bibitem{Kapusta} J.I. Kapusta and C, Gale, ``Finite Temperature Field Theory Principles and Applications," 2nd ed, Cambridge Univ. Press, Cambridge, UK, 2006.

\bibitem{Eric2} E. Braaten, Thermodynamics of Hot QCD, Nucl. Phys. {\bf A702}, 13 (2002).

\end{thebibliography}
\end{document}